\documentclass[aps,pra,reprint,letterpaper,amsmath,amssymb, superscriptaddress]{revtex4-1}
\usepackage{graphicx}
\usepackage[separate-uncertainty=true, per-mode=fraction]{siunitx}

\def\ge{g_{\text{e}}}
\def\mub{\mu_{\text{B}}}
\def\wl{\omega_{\text{L}}}
\def\wr{\omega_{\text{R}}}
\def\wn{\omega_{\text{N}}}

\begin{document}

\title{Decay and revival of electron spin polarization in an ensemble of (In,Ga)As quantum dots}

\author{E. Evers}
\affiliation{Experimentelle Physik 2, Technische Universit\"at Dortmund, 44221
Dortmund, Germany}
\author{V.~V. Belykh}
\affiliation{Experimentelle Physik 2, Technische Universit\"at Dortmund, 44221 Dortmund, Germany}
\author{N.~E. Kopteva}
\affiliation{Spin Optics Laboratory, St. Petersburg State University, 198504 St. Petersburg, Russia}
\author{I.~A. Yugova}
\affiliation{Spin Optics Laboratory, St. Petersburg State University, 198504 St. Petersburg, Russia}
\author{A. Greilich}
\affiliation{Experimentelle Physik 2, Technische Universit\"at Dortmund, 44221 Dortmund, Germany}
\author{D.~R. Yakovlev}
\affiliation{Experimentelle Physik 2, Technische Universit\"at Dortmund, 44221 Dortmund, Germany}
\affiliation{Ioffe Physical-Technical Institute, Russian Academy of Sciences, 194021 St. Petersburg, Russia}
\author{D. Reuter}
\affiliation{Department Physik, Universit\"at Paderborn, 33098 Paderborn, Germany}
\author{A.~D. Wieck}
\affiliation{Angewandte Festk\"orperphysik, Ruhr-Universit\"at Bochum, 44780 Bochum, Germany}
\author{M. Bayer}
\affiliation{Experimentelle Physik 2, Technische Universit\"at Dortmund, 44221 Dortmund, Germany}
\affiliation{Ioffe Physical-Technical Institute, Russian Academy of Sciences, 194021
St. Petersburg, Russia}

\begin{abstract}
The periodic optical orientation of electron spins in (In,Ga)As/GaAs quantum dots leads to the formation of electron spin precession modes about an external magnetic field which are resonant with the pumping periodicity.
As the electron spin is localized within a nuclear spin bath, its polarization imprints onto the spin polarization of the bath. The latter acts back on the electron spin polarization.
We implement a pulse protocol where a train of laser pulses is followed by a long, dark gap. It allows us to obtain a high-resolution precession mode spectrum from the free evolution of the electron spin polarization. Additionally, we vary the number of pump pulses in a train to investigate the build-up of the precession modes. To separate out nuclear effects, we suppress the nuclear polarization by using a radio-frequency field. We find that a long-living nuclear spin polarization imprinted by the periodic excitation significantly speeds up the buildup of the electron spin polarization and induces the formation of additional electron spin precession modes.
To interpret these findings, we extend an established dynamical nuclear polarization model to take into account optically detuned quantum dots for which nuclear spins activate additional electron spin precession modes.
\end{abstract}

\maketitle

\section{Introduction}
\label{sec:introduction}
Semiconductor quantum dots~(QDs) provide a convenient platform for the investigation and manipulation of the electron spin \cite{Loss1998}. One of the most important parameters related to the spin is its coherence time $T_2$ which is the time of phase loss for a spin precessing around a magnetic field. In view of practical applications, all manipulations on the electron spin based quantum bit have to be performed within the $T_2$ time. For an ensemble of QDs the measurement of $T_2$ is complicated due to inhomogeneities. The spread of the electron $g$ factors and effective magnetic fields of nuclear spins over QDs in an ensemble leads to dephasing of the spin polarization and fast decay of the average spin in the time $T_2^*$, typically $\sim 1$~ns \cite{Dutt2005,Petta2005,Greilich2006}, while spins in individual QDs preserve coherence on a microsecond time scale at liquid helium temperature \cite{Greilich2006a,Koppens2006,Koppens2008,Press2010,Bluhm2011,Bechtold2016,Stockill2016}. However, under certain conditions the above mentioned inhomogeneities can favor the measurement of $T_2$ and advanced spin manipulation. In Ref.~\cite{Greilich2006a}, it was found that pumping with periodic laser pulses leads to a selective spin orientation of the QDs with the Larmor spin precession frequency being a multiple of the pumping repetition rate. As a result of this spin mode locking (ML), the revival of the spin polarization occurs just before the arrival of the next laser pulse.
Furthermore, periodic pumping results in a polarization of nuclear spins bringing a majority of QDs in a ML condition through the hyperfine interaction known as the nuclei-induced frequency focusing (NIFF) effect \cite{Greilich2007,Dyakonov2017}.

Several mechanisms of NIFF were suggested \cite{Danon2008,Carter2009,Petrov2009,Korenev2011,Barnes2011,Glazov2012,Beugeling2016,Beugeling2017,Jaschke2017}. The major characteristics of the frequency focusing nuclear polarization is the distribution of the Overhauser field driving the spin precession of an electron in a QD into one of the resonant precession modes. The picture becomes even more involved due to the appearance of additional modes in the spectrum, pronounced at certain magnetic fields \cite{Jaschke2017}. So far, the electron spin precession frequency spectrum was not obtained directly in the experiments. Measurements were done under continuous periodic pumping which affects the spin system in a way similar to a periodic force shaking a pendulum whose dynamics would give only the frequency of the shaking force.

In this paper, using the recently developed extended pump-probe Faraday rotation technique \cite{Belykh2016}, we implement tailored pump pulse protocols and observe the free evolution of the QD spin ensemble after abruptly switching off the periodic pumping. The latter is analogous to the observation of free motion of a pendulum after release, which obviously gives the frequency and decay inherent to the system. But in contrast to the pendulum example, previous periodic pumping make imprints on the frequency spectrum of the QD spin ensemble. Furthermore, we are able to separate out the effect of NIFF on the frequency spectrum by applying a radio frequency~(RF) field covering all central and most quadrupole transitions which erases the nuclear polarization.

\section{Experimental details} \label{sec:experimentaldetails}

\begin{figure}[h]
   \includegraphics [width = \columnwidth]{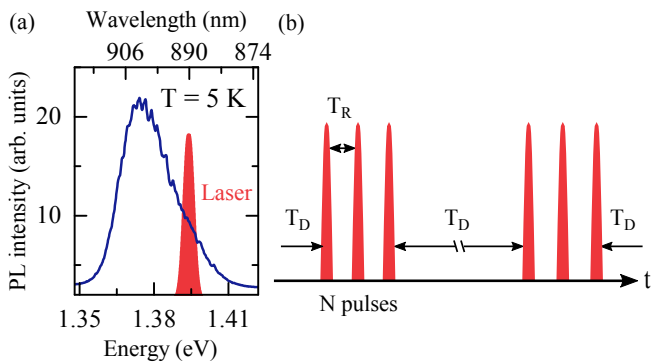}
   \caption{(Color online) Quantum dot excitation scheme. (a) Photoluminescence spectrum of the quantum dot ensemble at $T=5 \,$K and position of the laser energy ($1.393\,$eV). (b) The pump pulse sequence used in the experiment. The pulse trains are repeated quasi-infinitely with a dark gap of $T_{\text{D}}$ between them. Each pulse sequence consists of $N$ pulses, separated from one another by $T_{\text{R}}=13.2 \,$ns.}
\label{fig1}
\end{figure}

The studied sample consists of 20 layers of (In,Ga)As/GaAs QDs grown by molecular-beam epitaxy~(MBE) on a (001)-oriented GaAs substrate. The self-assembly of the QDs following the Stranski-Krastanov scheme leads to QDs consisting of $10^5$ atoms on average. To homogenize the dot size after growth, the sample was thermally annealed at $880 \, $\si{\degree C}, also shifting the central emission energy to about $1.38 \, $eV ($900 \, $nm).  The density of QDs in each layer is $10^{10} \, \text{cm}^{-2}$. The sample is $\delta$-doped by Silicon $16 \, $nm below each QD layer with a donor density similar to the QD density. It provides on average an ensemble of singly-charged QDs. Figure~\ref{fig1}(a) shows the photoluminescence (PL) spectrum of the QD ensemble. The sample is placed in a cryostat with a split-coil superconducting solenoid generating magnetic fields of up to $8 \, $T. The sample is brought into contact with helium gas at a temperature of $T=5 \, $K. A magnetic field is applied along the x axis perpendicular to the light propagation vector and the sample growth axis ($z$ axis), i.e., in Voigt geometry.

An extended pump-probe Faraday rotation technique is used to measure the electron spin dynamics at time delays ($1 \, \mu$s and more) strongly exceeding the laser repetition period using tailored pumping protocols while keeping a time resolution of about $2\,$ps~\cite{Belykh2016}. This technique represents a modification of the common pump-probe Faraday rotation technique, where circularly-polarized pump pulses generate a carrier spin polarization, which is then probed by the Faraday rotation of linearly-polarized probe pulses. The temporal evolution of the spin polarization is traced by varying the time delay between pump and probe pulses. In order to go for very long time delays and to have flexibility with setting excitation protocols we implement pulse picking for both pump and probe laser beams. In this way, a repeated train of a variable number of pump pulses can be applied to tailor a precession-mode structure. A graphic representation of an exemplary pulse sequence used in the experiment can be found in Fig.~\ref{fig1}(b).

\begin{figure*}
\includegraphics[trim=0mm 2mm 0mm 0mm, clip, width=2.05\columnwidth]{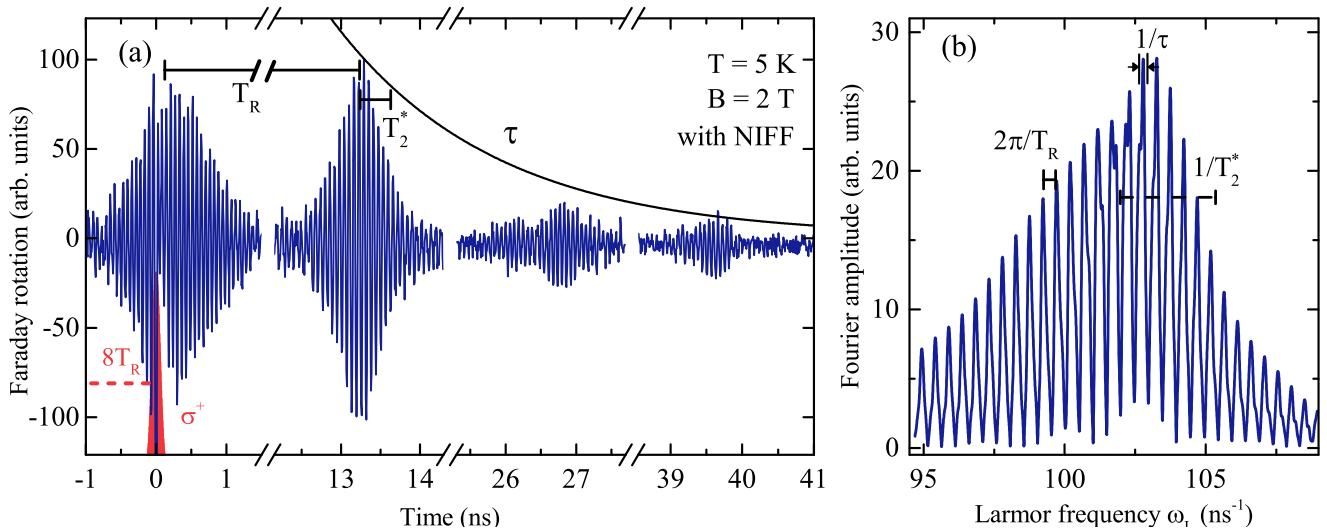}
\caption{(Color online) (a) Dynamics of the Faraday rotation signal after periodic application of a pulse train consisting of 8 pump pulses with a dark gap between trains of $1.6 \, \mu$s. Zero time corresponds to the arrival of the last pump pulse. (b) Spin precession spectrum obtained as the Fourier transform of the corresponding free evolution dynamics in (a) at $t>0$. $B=2 \,$T and $T=5 \,$K.}
\label{fig:Dyn}
\end{figure*}

We use a Ti:Sapphire laser emitting a train of $2\,$ps pulses with a repetition rate of $76\,$MHz (repetition period $T_\text{R}=13.2\,$ns) operating at $890\,$nm [in the high energy flank of the QD ensemble ground state emission, see Fig.~\ref{fig1}(a)]. The laser output is split into a pump and a probe beam. In the pump path an electro-optical modulator~(EOM) is installed to select sequences of $N$ pulses (from 1 to about 100) separated by $T_\text{R}$ with an arbitrarily long dark gap $T_\text{D}$ between the trains. By passing the EOM back and forth, the ratio of intensities of suppressed and transmitted pump pulses reaches about~$10^{-5}$. An acousto-optical light modulator~(AOM) in the probe path is used to select single pulses at the required delay after the pump train. Electronic variation of the delay between the synchronized AOM and EOM (also synchronized with the laser) in combination with a rather short mechanical delay line in the pump path allows for fine delay variation in a microsecond time range with $2\,$ps resolution. To perform synchronous detection, the polarization of the pump beam is modulated between $\sigma^{+}$ and $\sigma^{-}$ by a photo-elastic modulator (PEM) operating at a frequency of $84\,$kHz. Additionally, the probe beam is modulated between two linear polarizations by a PEM at 50 kHz. Using a linear polarizer we filter out only one polarization, which leads to the modulation of the probe intensity. The rotation of the transmitted probe polarization is then analyzed using a half-wave plate and a Wollaston prism followed by a balanced photo-diode scheme. Finally, a lock-in amplifier synchronized to the differential frequency of both pump and probe modulations is used to avoid any contribution of the scattered pump.

A small coil of about $1\,$mm core diameter serves to affect the nuclear spins. The coil is mounted directly in front of the sample to induce an RF field along the optical axis. The coil is connected via a $10\,$dB attenuator to an RF amplifier, which is driven by a function generator. We do not implement an impedance matching circuit in order to apply a spectrally broad RF field in a frequency range from $13$ to $28 \, $MHz which covers the spin precession frequencies of the nuclear isotopes of In, Ga, As at a magnetic field of $2 \,$T. The RF field is periodically present for $1\,$ms and absent for $10\,$ms due to the duty cycle of the amplifier. Any direct effect of the RF field on the electron spins would thus only contribute by less than $10\,$\% to the overall Faraday rotation signal. An alternating magnetic field of up to $8\,$mT along the $z$ axis could be achieved this way.

\section{Experimental results} \label{sec:results}

\begin{figure}
  \includegraphics[width=0.98\columnwidth]{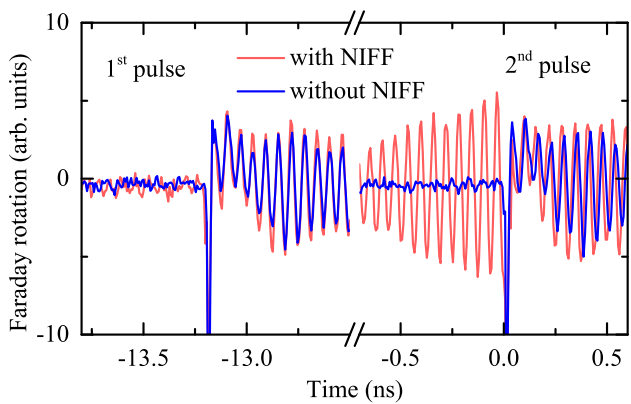}
  \caption{(Color online) Dynamics of the Faraday rotation signal for a train of two pulses with a dark gap between pulses of $1.7 \, \mu$s. Zero time corresponds to the arrival of the second pulse. The red line and the blue line correspond to the RF field off and on, respectively. $B = 2\, $T and $T = 5\, $K.} \label{fig:1and2}
\end{figure}

Figure~\ref{fig:Dyn}(a) shows the dynamics of the Faraday rotation signal at $B=2 \, $T after a train of 8 pump pulses, where zero time ($t=0$) corresponds to the arrival of the last pump pulse. The frequency of the oscillatory part corresponds to the average electron Larmor frequency $\wl$
  \begin{align}
  \wl = \frac{|\ge| \mub B}{\hbar}
  \label{eq:wl}
  \end{align}
with the electron $g$-factor $\ge$, the Bohr magneton $\mub$ and the total magnetic field $B$. Starting from $t=0$, the signal decays within an inhomogeneous dephasing time $T_2^* \approx 0.5$~ns. The dephasing is the result of a non-uniform Larmor frequency distribution in the electron spin ensemble due to the spread of the electron $g$ factors in the QD ensemble. However, around every $13.2\,$ns (the repetition period of the laser) bursts of the signal occur due to the rephasing of the electron spin polarization. The same effect is responsible for the increase of the electron spin polarization at negative delay, before the arrival of the last pump pulse. This is referred to as spin mode locking~\cite{Greilich2006a}. The amplitude of the polarization bursts rapidly decreases with the burst number, so the fourth burst decays below the sensitivity limit.
The measured Faraday rotation dynamics for $t > 0$ gives the free evolution of the average electron spin polarization in the ensemble. The Fourier transform of this dynamics gives the spectrum of the spin precession modes in the ensemble of electron spins which is presented in Fig.~\ref{fig:Dyn}(b). The spectrum has a finite half width at half maximum of about $3\,$ns$^{-1}$ determined by the inhomogeneous dephasing time $T_2^*$. It consists of well pronounced modes with a separation of $2\pi/T_\text{R}$, corresponding to the periodic bursts of the signal. The width and the shape of the modes reflect the bursts' decay.

\begin{figure*}
  \includegraphics[width=2.05\columnwidth]{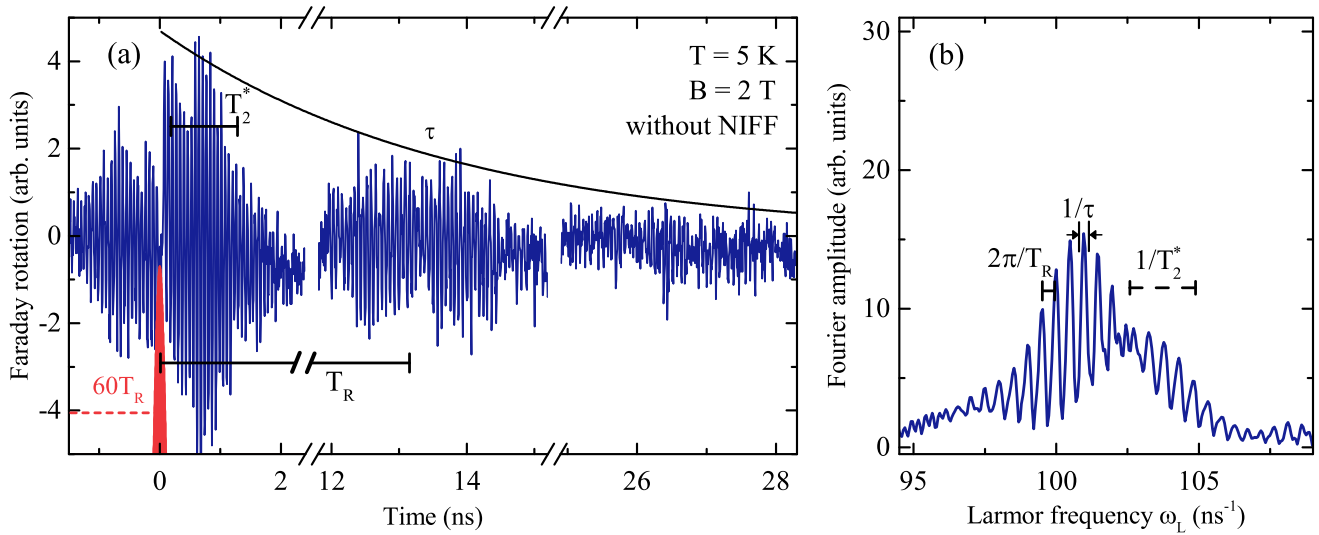}
  \caption{(Color online) Dynamics of the Faraday rotation signal after a train of 60 pump pulses with a dark gap of $1.6 \, \mu$s when the nuclei are depolarized by an RF field. Zero time corresponds to the arrival of the last pump pulse. (b) Respective Fourier transform spectrum of the dynamics at $t>0$. $B=2$~T and $T=5$~K.} \label{fig:DynRF}
\end{figure*}

The spin precession frequency spectrum [Fig.~\ref{fig:Dyn}(b)] is contributed by the electron $g$-factor distribution, by the interaction with nuclei, and by the excitation conditions. In particular, the photon energy of the exciting laser determines the average $g$ factor of QDs whose electrons acquire spin polarization \cite{Schwan2011a,Belykh2016a}. The periodicity of the laser excitation implies efficient polarization of the QDs with electron spin frequencies satisfying the relation
\begin{equation}
\omega_{\text{L}} = n \wr = \frac{2 \pi n }{T_\text{R}},
\label{eq:ML}
\end{equation}
where $n$ is an integer number. This is the essence of the ML effect \cite{Greilich2006a}. However, even for QDs having precession frequencies that do not satisfy Eq.~\eqref{eq:ML}, the effective nuclear magnetic field $B_\text{N,x}$ (Overhauser field) is adjusted in a way that it contributes to the total magnetic field and makes the electron spin precession frequency satisfying relation \eqref{eq:ML}. Only the Overhauser field in direction of the external magnetic field $B_{\text{N},z}$ significantly modifies each electron spin Larmor precession. This is the nuclei-induced frequency focusing (NIFF) effect which enhances the amplitude of the ML. The nuclear polarization introduced in this way persists for tens of minutes without further electron excitation and for several seconds when actively pumping the electron-spin polarization and forcing another precession-mode structure~\cite{Greilich2007}. Note, that in the present theories of the ML and NIFF effects, the mode width is mainly determined by the homogeneous electron spin relaxation time $T_2$, which in the studied (In,Ga)As/GaAs QDs is $\sim 1 \, \mu$s~\cite{Greilich2009}. This fact, however, was not experimentally verified so far. As we see now from the experiment the mode width is much larger than $1/T_2$ or equivalently the bursts' decay is much faster than $T_2$.

In what follows we suppress the contribution of the NIFF to the ML by depolarizing nuclei with an RF field. The depolarization of the nuclear spins by the RF field is demonstrated in the Appendix.
Figure~\ref{fig:1and2} shows the effect of the NIFF on the ML. In this experiment, we pump the electron spins with repeating trains of 2 pulses. The pulses are separated by $T_\text{R}$. The separation between the trains was $T_{\text{D}}=128 T_\text{R} \approx 1.7 \, \mu$s. Note that $T_{\text{D}}$ is longer than $T_{2}$. Before the first pulse no ML is observed and the signal after the pulse is the same with and without RF. Thus, RF has no effect on the electron spin polarization created by a single pulse. Before the second pulse still no ML is observed when RF is switched on. Indeed, the electronic system looses its coherence between the pump pulse trains and correspondingly loses the information about the arrival of the second pulse. In contrast, when not applying the RF field, a mode-locking signal appears before the second pulse. We attribute its appearance to the specific nuclear Overhauser field distribution, emerged due to the repeated application of the pulse trains, which superimposes the external magnetic field. Electron spins subject to this Overhauser field precess with frequencies satisfying the ML relation \eqref{eq:ML} in a way that leads to the formation of a significant ML signal. The information on the pump-pulse sequence is hence stored in the nuclear spin Overhauser field. We conclude that the application of the RF field induces random nuclear spin flips leading to its destruction as if no nuclear polarization has taken place.

The dynamics of the Faraday rotation signal after a train of 60 pump pulses with applied RF field is shown in Fig.~\ref{fig:DynRF}(a). In this case, the NIFF is absent and the ML is the result of preferential polarization of the QDs satisfying Eq.~\eqref{eq:ML} only \cite{Greilich2006a}. The Fourier spectrum of the dynamics after the last pump pulse is shown in Fig.~\ref{fig:DynRF}(b). It shows smaller modulation amplitude compared to the spectrum without RF [Fig.~\ref{fig:Dyn}(b)] which is expected since the NIFF makes modes more pronounced by forcing a larger number of electron spins to satisfy the ML condition. Take note of the fine structure of the modes, i.e. the deviation of the mode shape from a symmetric peak, which is present when the RF is off [Fig.~\ref{fig:Dyn}(b)] and almost absent when the RF is on [Fig.~\ref{fig:DynRF}(b)]. This fine structure is presumably related to the Overhauser (nuclear) field distribution imprinted by the periodic pumping~\cite{Beugeling2016,Beugeling2017,Jaschke2017}.

\begin{figure}[h]
  \includegraphics[width=0.98\columnwidth]{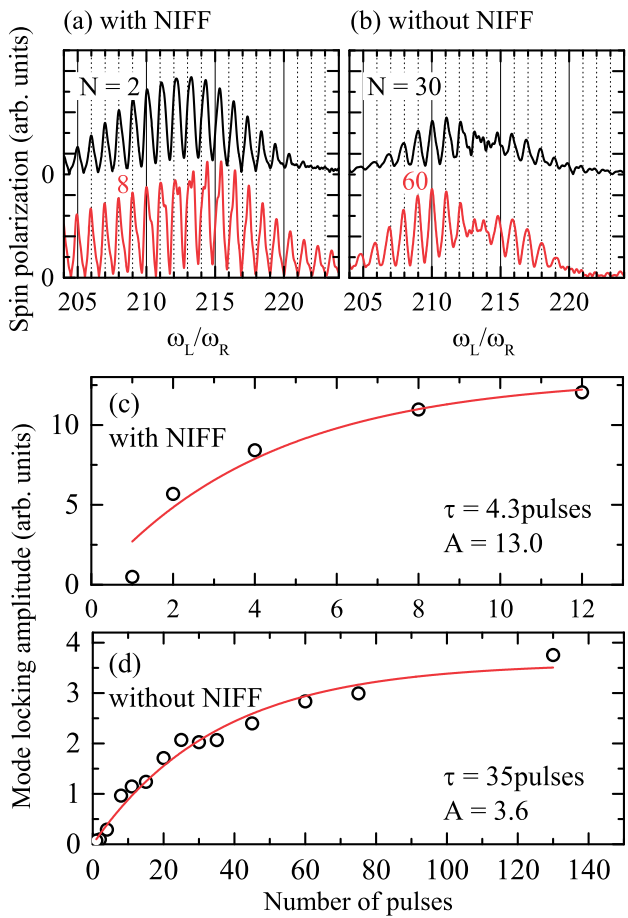}
  \caption{(Color online) (a),(b) Electron spin precession spectra for different number of pulses in a pump train with (a) and without (b) nuclei-induced frequency focusing~(NIFF) controlled by the application of the RF field. The Fourier spectra are calculated by transforming the free dynamics from the last pump pulse. (c),(d) Dependence of the mode locking amplitude (amplitude of the first burst) on the number of preceding pulses in a train with NIFF (c) and without NIFF (d). $B = 2\, $T and $T = 5\, $K.} \label{fig:nffbuildup}
\end{figure}

Furthermore, we register spin polarization bursts for an increasing number of pulses in a pump train. In this way, the formation of the respective mode-structure can be traced. Figures~\ref{fig:nffbuildup}(a) and \ref{fig:nffbuildup}(b) show the electron spin precession spectra for a different number of pulses in a pump train with (a) and without (b) NIFF controlled by the application of the RF field. When the RF field is absent and the electron spin precession is shaped by NIFF [Fig.~\ref{fig:nffbuildup}(a)], the amplitude of the central modes for 2 pulses is as high as the amplitude after 8 pulses. The difference lies here in the formation of side modes which appear only with an increasing number of pump pulses. Moreover, the modulation depth also increases. A much higher number of pump pulses is needed for the spin precession buildup when the NIFF is destroyed by the RF field [Fig.~\ref{fig:nffbuildup}(b)]. In this case, an increase in the number of pump pulses primarily leads to an increase in the amplitude of the spin precession modes, while the number of active modes is barely affected.

Finally, we summarize the dynamics of the spin polarization with increasing number of pump pulses. Figures~\ref{fig:nffbuildup}(c) and \ref{fig:nffbuildup}(d) show the amplitude of the ML signal (the amplitude of the first burst) as a function of the number of pulses in a train with and without NIFF i.e. without and with RF field, respectively. The spin polarization is fitted by an exponential saturation law $y=A\left[1-\exp(- N/N_\text{sat}) \right]$ to obtain the characteristic number of pulses $N_\text{sat}$ required for the ML to set in. It is seen from the figures that the NIFF effect strongly enhances the ML which sets in after $\sim 4$ pump pulses. Here, as it is shown above, the increase in the ML amplitude mainly results from the increase in the number of the spin precession modes. On the other hand, without NIFF, when the RF field is on, the onset of ML is slow and requires $\sim 35$ pulses. It results mainly from the increase in the amplitude of the spin precession modes. Furthermore, the maximal ML amplitude with RF is $\sim 4$~times lower than that without RF.

\section{Theory}
Here, we present a model that treats the process of the electron spin polarization and its dynamics accounting for the NIFF effect.

\subsection{Electron spin polarization}
We assume that the spin coherence of resident electrons is generated via trion excitation. One can express the spin polarization of individual resident electrons localized in QDs ${\bf S}^a$ after an infinite sequence of pump pulse trains as~\cite{Yugova2009,Yugova2012}:
\begin{eqnarray}
{\bf S}^a =  I[I-[\mathcal A\cdot \mathcal B(T_\text{R})]^{N-1}\mathcal A\cdot \mathcal B (T_\text{D})]^{-1} \\ \nonumber
I[I-[\mathcal A\cdot \mathcal B(T_\text{R})]^{N-1}][I - \mathcal A\cdot \mathcal B(T_\text{R})]^{-1} \cdot {\bf S}_0. \label{eq:sa}
\end{eqnarray}
Here $I$ is the unity matrix. We remind that $T_\text{R}$ is the pump pulse repetition time, $T_\text{D}$ is the time between the last pump pulse in a train and the first pulse in the next train, and $N$ is the number of pump pulses in a train [see Fig. \ref{fig1}(b)]. Matrices $\mathcal A$ and $\mathcal B$ describe optical spin orientation by a pump pulse and the spin dynamics in the magnetic field, respectively:

\begin{equation}
\mathcal A = \begin{pmatrix}
Q \cos \Phi & Q \sin \Phi & 0\\
- Q \sin \Phi & Q \cos \Phi & 0\\
0 & 0 & \frac{Q^2 + 1}{2}
\end{pmatrix},
\mathcal {\bf S}_0 = \begin{pmatrix}
0 \\
0 \\
\frac{Q^2 - 1}{4}
\end{pmatrix},
\end{equation}
\begin{equation}
\mathcal B(t) = \begin{pmatrix}
1 & 0  & 0\\
0 & \cos \omega_\text{L} t & - \sin \omega_\text{L} t\\
0 & \sin \omega_\text{L} t & \cos \omega_\text{L} t
\end{pmatrix}  \exp( {-\frac{t}{T_2}}) .
\end{equation}

\begin{figure}[h]
  \includegraphics [width=0.9\columnwidth]{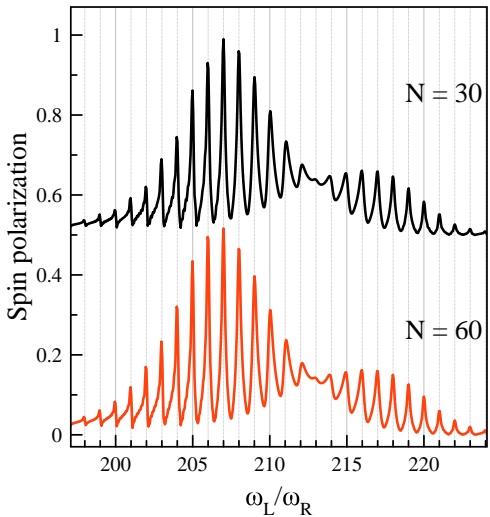}
  \caption{(Color online) Calculated spin precession spectra after different numbers $N$ of pump pulses in a train without nuclear effects. The curves are vertically shifted for clarity. The parameters of the calculations: $T_2 = 45T_\text{R}$, $T_2 < T_\text{D}$, $T_\text{D} = 130T_\text{R} - NT_\text{R}$, and pulse area $\Theta = \pi$.}
  \label{fig2}
\end{figure}

The Larmor precession frequency $\omega_\text{L}$ is defined by Eq.~\eqref{eq:wl}, $t$ is the time delay, $T_2$ is the electron spin coherence time, $Q$ is the pump pulse amplitude, and $\Phi$ is the pulse phase (see equations for Rosen-Zener pulse (22-23) in \cite{Yugova2009}):
\begin{equation}
Q = \left| \frac{\Gamma^2(\frac{1}{2}-i\Delta)}{\Gamma(\frac{1}{2}-i\Delta-\frac{\Theta}{2\pi})\Gamma(\frac{1}{2}-i\Delta+\frac{\Theta}{2\pi})}\right|
\end{equation}
\begin{equation}
\Phi = \arg \left(\frac{\Gamma^2(\frac{1}{2}-i\Delta)}{\Gamma(\frac{1}{2}-i\Delta-\frac{\Theta}{2\pi})\Gamma(\frac{1}{2}-i\Delta+\frac{\Theta}{2\pi})}\right)
\end{equation}
here $\Delta=(E_\text{p}-E)\tau_\text{p}/2\pi\hbar$ is the optical detuning of the QD transition energy $E$ with respect to the pump photon energy $E_\text{p}$, $\Theta = \mu \tau_\text{p}$ is the pulse area, the coefficient $\mu$ is the measure proportional to the electric field strength, and $\tau_\text{p}$ is the pulse duration time. We take into account that $T_2 \gg \tau_\text{p}$ as well as $1/\wl \gg \tau_\text{p}$.

The experimentally detected Faraday rotation signal can be described as
\begin{eqnarray}
\mathcal F & = & \int \left[S^a_z(\omega_\text{L},E)\rho_\text{tr}(\omega_\text{L},E)G_\text{tr}(E_\text{p},E) \right]dE.
\label{eqF}
\end{eqnarray}
Here $\rho_\text{tr}(\omega_\text{L},E)$ is the distribution function of the Larmor precession frequencies for the QDs having an optical transition energy $E$. For simplicity we assume a single-valued dependence between the electron $g$ factor (and, thus, $\omega_\text{L}$) and $E$, similarly to the case for bulk semiconductors \cite{Roth1959}. In this case $\rho_\text{tr}(\omega_\text{L},E)$ are $\delta$ functions. Furthermore, for small variation of $E$ we assume a linear dependence $|\ge| = \bar A E + \bar C$ and enhance the constant $\bar A$ to account for an additional spread of $g$ factors due to the dependences on the QD shape and composition. For calculations $\bar A = -2.5$ \, eV$^{-1}$ and $\bar C = 3.0$ were used. Due to $\bar A < 0$ the absolute value of the electron $g$ factor decreases for increasing optical energy. Hence, negatively detuned electrons feature smaller mode numbers $\omega / \wr $, while positively detuned electrons feature lager mode numbers.

The function $G_\text{tr}(E_\text{p},E) = G_\text{tr}(\Delta)$ describes the spectral dependence of the trion Faraday rotation signals on the optical detuning $\Delta$. We take $G_\text{tr}(\Delta)$ as the imaginary part of the generalized Riemann function \cite{Yugova2009}:
\begin{equation}
G(\Delta) = \frac{\tau_\text{p}^2}{\pi^2}\sum_{K=1}^{\infty}(K+\frac{1}{2}-i\Delta)^{-2},
\end{equation}
where $K$ is integer.

The calculated spin precession spectra are shown in Fig.~\ref{fig2} for a different number of pulses in a train. The mode structure originates from the resident electrons' spin polarization after a train of pulses. The spectrum shows two wings of polarization around mode number $213$. This feature stems from the spectral behavior of the Faraday rotation. The asymmetrical shape of the wings originates from the spectral distribution of QDs [Fig. \ref{fig1}(b)]. Note that the spin precession spectrum calculated without nuclear contribution is not broadened with increasing number of pulses (Fig.~\ref{fig2}) in agreement with the experiment [Fig.~\ref{fig:nffbuildup}(b)].

\subsection{Electron-nuclei spin dynamics}
As a next step, we take into account the effect of nuclear polarization in our model. In case of a nonzero optical detuning $\Delta \neq 0$ and a pulse phase $\Phi \neq 0$ the pump pulses generate a nonzero $x$-component of the average electron spin polarization $\langle S_x\rangle$ (note, $\mathbf{B}||\mathbf{x}$) \cite{Carter2009}. The electrons try to reach their thermal equilibrium populations and, thus, provide energy for a spin-flip process which results in an increase of the nuclear polarization. This nuclear polarization can be considered as an additional magnetic field (the Overhauser field). The feedback of the nuclear spin system on the electron spin changes the Larmor frequency of the electron spin precession \cite{Carter2009,Carter2011,Korenev2011}.

\begin{figure}[h]
  \includegraphics [width = 0.98\columnwidth]{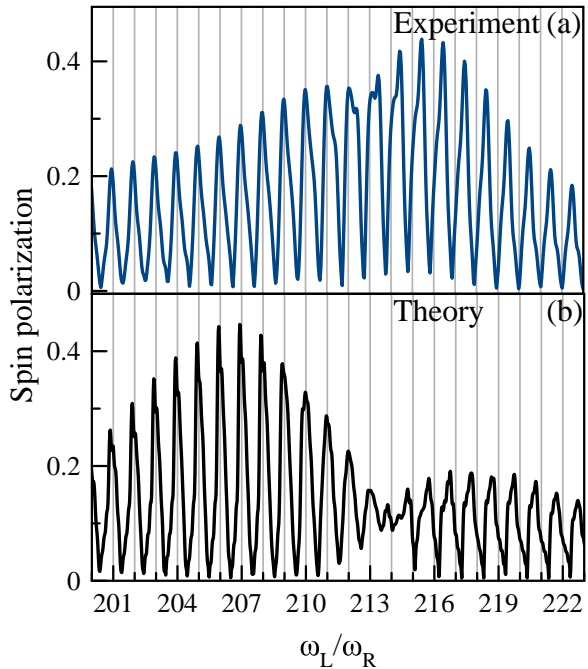}
  \caption{(Color online) Comparison of the measured (Fig. \ref{fig:Dyn}) (a) and calculated (b) electron spin precession spectra with NIFF.
    Due to the interaction with nuclei, the spin precession modes for the high-frequency QDs are located in between multiple integers of the repetition frequency. The parameters of the calculations: $T_2 = 45T_\text{R}$, $\Theta = \pi$, $B = 2$~T, $\chi = 0.39$, $A_\text{N} = 48.5 \mu eV$, $\bar{q} = 5$, $f_\text{N} = 0.05$, $T_\text{D} = 130 T_\text{R} -N T_\text{R}$, and $N = 8$.}
  \label{fig3}
\end{figure}

The dynamic nuclear polarization (DNP) model can be used when the nuclear Overhauser field fluctuations are smaller than the averaged Overhauser field. The averaged nuclear spin $I_\text{N}$ and Overhauser field $B_\text{N}$ are calculated using the following equations \cite{Dyakonov1974}:
\begin{equation}\label{eq:IN}
\frac{dI_\text{N}}{dt} = - \frac{1}{T_\text{1e}} [I_\text{N} - \bar{q} \langle S_{x}(I_\text{N})\rangle] - \frac {I_\text{N}}{T_{1L}},
\end{equation}
\begin{equation}\label{eq:BN}
B_\text{N} = \frac{A_\text{N} I_\text{N} \chi}{\mu_\text{B} |\ge|},
\end{equation}
where $T_\text{1e}$ is the nuclear spin relaxation time due to interaction with electrons and $T_\text{1L}$ is the nuclear spin-lattice relaxation time, which also takes into account any other possible relaxation processes not related to the interaction with the electron, $\bar{q} = 4 \bar{I}(\bar{I}+1)/3$. For simplicity we take only one type of nuclear with spin $\bar{I} = 3/2$, corresponding to $^{71}\text{Ga}$. Here, $A_\text{N}$ is the hyperfine constant and $\chi$ is the nuclear isotope abundance. We first calculate the average electron spin component $S_x$ without nuclear contribution. Using Eqs.~(\ref{eq:IN}-\ref{eq:BN}) we calculate the Overhauser field which is then used to recalculate the electron spin dynamics. The Overhauser field causes a shift of the electron precession frequency. For the Overhauser field calculations, we use the following parameters: $A_\text{N} = 48.5 \mu eV$, $\bar{I} = 3/2$, $\bar{q} = 5$, $\chi = 0.39$, and the leakage factor $f_\text{N} = T_\text{1L}/(T_\text{1L}+T_\text{1e}) = 0.05$. The calculated electron spin precession spectrum is shown in Fig.~\ref{fig3}(b) in comparison to the experimental one [Fig.~\ref{fig3}(a)].

\begin{figure}[h]
  \includegraphics [width = 0.98\columnwidth]{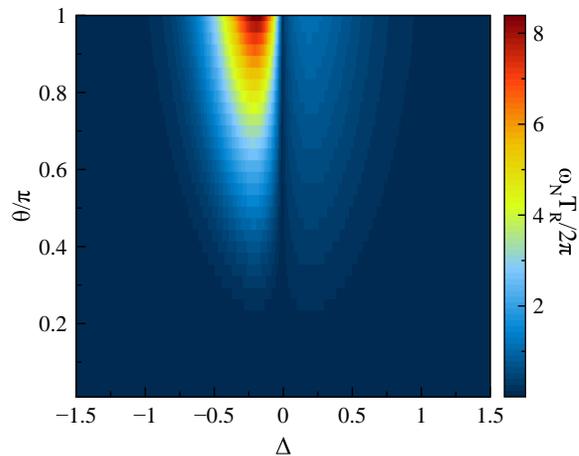}
  \caption{(Color online) Amplitude of the Overhauser field in units of the mode number $\omega_{\text{N}} T_{\text{R}}/2\pi$ as a function of the detuning $\Delta$ ($x$-axis) and the pump pulse area $\Theta$ ($y$-axis). The parameters of the calculations: infinite number of pump pulses; $T_2 = 7T_\text{R}$, $\chi = 0.39$, $A_\text{N} = 48.5 \mu eV$, $\bar{q} = 5$, and $f_\text{N} = 0.1$.}
  \label{fig5}
\end{figure}

It is important to note that in the case of negatively detuned QDs, when $E>E_\text{p}$ (smaller precession frequencies in Fig.~\ref{fig3}), the Overhauser field tends to move the electron precession frequency towards the phase synchronization conditions. For positively detuned QDs (larger frequencies in Fig.~\ref{fig3}), the nuclear polarization changes to move the electron precession frequency away from the phase synchronization conditions~\cite{Carter2009,Carter2011}. This results in a shift of the mode positions out of condition \eqref{eq:ML} which is also observed in the experiment [Fig.~\ref{fig3}(a)]. The experimental spectrum and the modeled spectrum differ. The position of zero detuning around mode number $213$ is not pronounced in the experimental data. For the model calculations, we only use a small ensemble of electron spins without a nuclei-induced shifting of the modes by an amount larger than a single mode number. In the experiment, however, we probe a large, asymmetric ensemble whose precession modes might be shifted significantly and the overlap of different ensembles leads to an overlap of modes on integer numbers and in between integer numbers. Moreover, the asymmetry present in the model results and in the data without NIFF [Fig.~\ref{fig:nffbuildup}(b)] is not preserved in the experimental spectrum with NIFF. In the experiment, we observe that the positively detuned modes (in between integer numbers) are polarized more efficiently.

As it was mentioned, the precession spectrum without nuclear contribution does not broaden with an increasing number of pulses. The situation dramatically changes when the NIFF effect is due [Fig.~\ref{fig:nffbuildup}(a)]. The number of electron spin precession modes is increasing and the spectrum becomes broader with nuclear influence. This is related to the Overhauser nuclear field, which overlaps several modes of synchronization.

We explore how the Overhauser field depends on the parameters of the QD excitation: The pulse area $\Theta$ and the detuning $\Delta$ between the pump photon energy and the QD transition energy. For this we consider a model situation of an infinite number of pulses in a train. In this case, equation~\eqref{eqF} allows to find an analytical expression of the $x$-component of the electron spin polarization (Eqs.~(29,30) in~\cite{Yugova2009}).

The dependence of the absolute value of the estimated Overhauser field on $\Theta$ and $\Delta$ is shown in Fig.~\ref{fig5}. The Overhauser field is given as the number of modes which this field can overlap: $\wn T_\text{R} /2\pi$, where \linebreak $\wn = |\ge| \mu_\text{B} B_\text{N} / \hbar$. The overlap increases to up to 8 modes with increasing pump intensity, as $\Theta/\pi$ approaches to 1. In this way, the electron Larmor frequency may be shifted by an amount $\wn \sim 8 \times 2\pi/T_\text{R}$ out of the initial width of the spectrum leading to its broadening. Increasing the number of pulses in the train leads to an increase in the Overhauser field broadening the spectrum. Note that the Overhauser field distribution is asymmetrical. It means that in the case of increasing number of pulses in the train we should observe an asymmetrical increase in the number of the electron precession modes.

\section{Conclusions}
In conclusion, we experimentally reveal the electron spin precession spectra in quantum dots pumped by trains of laser pulses. We investigate the contribution of the nuclear spin polarization to the electron spin dynamics by applying a radio-frequency field which suppresses the nuclear spin polarization. The electron spin precession forms a mode structure dictated by the pumping periodicity. The mode width is much larger than $1/T_2$, which is unexpected in the framework of the existing spin mode locking theories. The presence of the nuclei-induced frequency focusing makes modes much more pronounced and induces additional modes at the positions of the mode locking antiresonances. We explain this effect within a dynamical nuclear polarization model by the differences in the nuclear focusing mechanism for positively and negatively detuned QDs. It was found that an increase in the number of  pulses in the pumping train leads to an abrupt saturation of the central modes and a fast broadening of the spin precession spectrum, i.e. the Overhauser field pushes electron spins in side modes that are not pumped directly. Without nuclear contribution, the increase in the number of pump pulses leads to a slow increase in the modes amplitude only. The significant difference in the electron spin polarization buildup rates with and without nuclear frequency focusing stands to be explained and probably needs taking into account fluctuations of the Overhauser field \cite{Danon2008} acting on a submicrosecond timescale \cite{Bechtold2015}.

\begin{acknowledgments}
  We are grateful to F.B. Anders, W. Beugeling, and G. Uhrig for useful
  discussions. This work was supported by Deutsche Forschungsgemeinschaft and the Russian Foundation of Basic Research (Grant No. 15-52-12019NNIOa) in the frame of ICRC TRR 160 (Project A1). N.E.K. and I.A.Yu. acknowledge support from Saint-Petersburg State University (Grant No. 11.34.2.2012).
\end{acknowledgments}

\section{Appendix}

\begin{figure}
  \includegraphics[width=0.98\columnwidth]{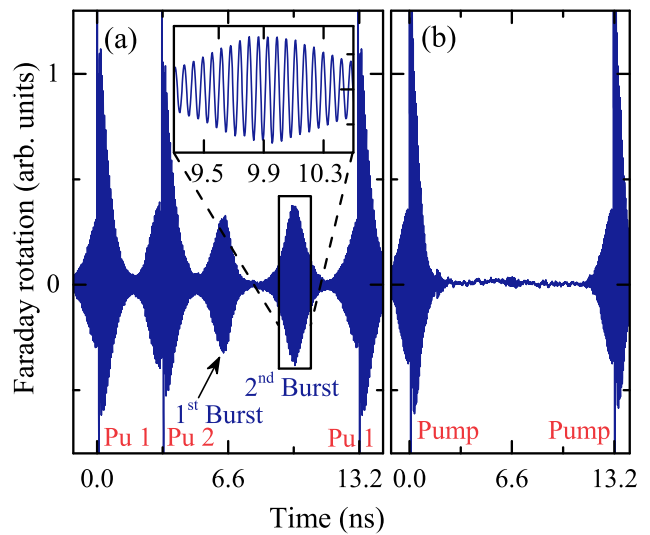}
  \caption{(Color online) Formation of electron spin precession bursts under resonant pumping. (a) A second pump pulse~($\pi$-pulse) implements boundary conditions which force the emergence of spin polarization bursts. (b) In a pulse sequence without a second pump pulse, spin polarization bursts only manifest as spin mode locking before the next pump pulse. $B = 2\, $T at $T = 5\, $K.} \label{fig:Bursts}
\end{figure}

\begin{figure}
  \includegraphics[width=0.98\columnwidth]{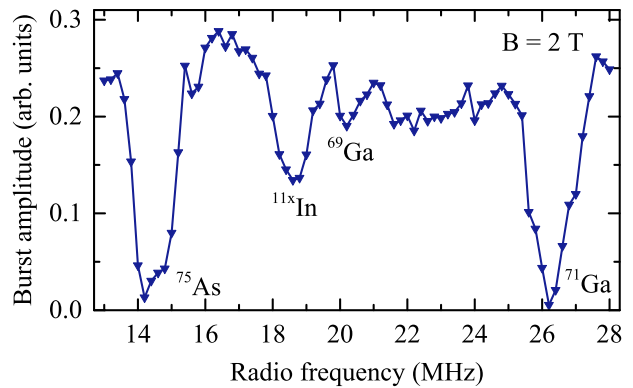}
  \caption{(Color online) Amplitude of the second electron spin polarization burst after switching off the second laser pulse versus RF field frequency. All nuclear spins are depolarized at their predicted nuclear magnetic resonance frequencies. $B = 2\, $T at $T = 5\, $K.} \label{fig:NuclSpectrum}
\end{figure}

In order to manipulate the nuclear spins via an RF field, we identify the nuclear spin resonance frequencies. The nuclear (Overhauser) field, that is established through nuclei-induced frequency focusing (NIFF), forces electron-spin precession frequencies to satisfy the mode-locking condition imprinted by a pump pulse pattern (see Fig.~\ref{fig:Bursts}). Here, we initially implement an experimental configuration where a continuous sequence of pump-pulse pairs was used. The repetition period between pairs amounts to $T_\text{R} = 13.2 \,$ns, while the separation between pulses within a pair is $T_\text{R}/4$. After the second pump pulse in a pair, spin polarization bursts with repetition period $T_\text{R}/4$ are observed. The bursts persist for seconds after the second pump pulse is switched off. Then, the presence of the bursts is a clear indicator for a remaining nuclear polarization. Subsequent application of an RF field can decrease the amplitude of the bursts when its frequency falls in resonance with the spin precession frequencies of nuclear isotopes present in the QDs as shown in Fig.~\ref{fig:NuclSpectrum}. Thus, we demonstrate that the RF field depolarizes nuclear spins and destroys the Overhauser field distribution imprinted by the pumping sequence. In the experiments described in the main part of the present work, we apply long RF pulses whose frequencies' cover all the nuclear isotopes resonances in the QDs. The amplitude of the RF field was chosen in the saturation region, when a further increase of this amplitude has no effect on the mode locking signal.

\bibliography{QD-Decay-and-Revival_ExtPuPr_EEvers}

\end{document}